\documentclass[aps,pra,twocolumn,oneside,showpacs,superscriptaddress,nofootinbib]{revtex4-2}
\usepackage{graphicx}  
\usepackage{dcolumn}   
\usepackage{placeins}
\usepackage{tikz}
\usepackage{lipsum}
\usepackage{soul}
\usepackage{float}
\makeatletter
\let\newfloat\newfloat@ltx
\makeatother
\usepackage{algorithm}
\usepackage{algorithmic}
\makeatletter
\renewcommand{\fnum@algorithm}{Algorithm~\thealgorithm}
\makeatother
\usepackage{physics}
\usepackage{siunitx}
\usepackage{ORCIDinREVTeX}

\usepackage{graphicx}
\usepackage{amsmath}
\usepackage{amsfonts}
\usepackage{epsfig}
\usepackage{slashed}
\usepackage{braket}
\usepackage{comment}
\usepackage{cases}
\usepackage{appendix}
\usepackage{color}
\usepackage{booktabs}
\usepackage{lipsum}
\definecolor{mygreen}{rgb}{0,0.5,0}

\def\bec{\begin{center}}
\def\eec{\end{center}}
\def\beq{\begin{equation}}
\def\eeq{\end{equation}}
\def\bea{\begin{eqnarray}}
\def\eea{\end{eqnarray}}

\usepackage{lineno}
\usepackage[unicode=true,
            pdfusetitle,
            colorlinks=true,
            urlcolor=black,
            anchorcolor=black,
            citecolor=red,
            filecolor=navyblue,
            linkcolor=red,
            menucolor=navyblue,
            linktocpage=true,
            pdfproducer=medialab]{hyperref}
\hypersetup{pdfauthor={Name}}

\begin{document}
\title{Information Propagation in Rydberg Arrays via Analog OTOC Calculations}

\author{Goksu Can Toga}
\orcid{0000-0002-0316-2502}
\email{gctoga@ncsu.edu}
\affiliation{Department of Physics and Astronomy, North Carolina State University, Raleigh, NC 27695, USA }

\author{Siva Darbha}
\affiliation{National Energy Research Scientific Computing Center, Lawrence Berkeley National Laboratory, Berkeley, CA 94720, USA}

\author{Ermal Rrapaj}
\affiliation{National Energy Research Scientific Computing Center, Lawrence Berkeley National Laboratory, Berkeley, CA 94720, USA}

\author{Pedro L. S. Lopes}
\affiliation{QuEra Computing Inc., 1284 Soldiers Field Road, Boston, MA 02135, USA}

\author{Alexander F. Kemper}
\orcid{0000-0002-5426-5181}
\email{akemper@ncsu.edu}
\affiliation{Department of Physics and Astronomy, North Carolina State University, Raleigh, NC 27695, USA }

\date{\today}

\begin{abstract}
Out-of-time-order correlators (OTOCs) are the main tool for probing quantum chaos and scrambling, and have become crucial probes in many areas of quantum computing. However, the measurement of OTOCs is difficult to implement on analog quantum computers due to the requirement of backward time evolution. In this paper, we develop and implement a randomized measurement protocol to compute OTOCs on Aquila by QuEra Computing. Unlike traditional methods that require backward time evolution, our approach utilizes a sequence of global randomized quenches that approximates the unitary 2-design properties necessary for extracting infinite-temperature OTOCs from statistical correlations. We demonstrate the protocol's success by explicitly observing the lightcone of information propagation in 1D Rydberg chains, and compare hardware results to both state-vector simulations and matrix product state (MPS) tensor network calculations. This work establishes the first demonstration of fully analog randomized OTOC measurements in neutral-atom simulators, providing a scalable pathway to probe quantum chaos in complex many-body systems.

\end{abstract}

\maketitle

\section{Introduction}

Out-of-time ordered correlators (OTOCs) have become an essential tool for characterizing quantum many-body systems, since they probe information propagation, scrambling dynamics, and quantum chaos~\cite{sekino_fast_2008,Maldacena:2015waa,swingle2016measuring,xu_accessing_2020,swingle2018unscrambling,xu2024scrambling,toga2025fast,dowling2023scrambling,xu2020does,hashimoto2020exponential,pappalardi2018scrambling,Shenker:2013pqa}. While they were first introduced to study quasiclassical superconductivity~\cite{larkin1969quasiclassical}, more recently they have become important diagnostics for rapidly advancing quantum computing and simulation platforms used in various applications, from diagnosing quantum chaos to Hamiltonian learning~\cite{li_measuring_2017,garttner2017measuring,joshi2020quantum,mi2021information,landsman2019verified,abanin2025constructive,zhang_direct_2009,doi:10.1073/pnas.2321668121,chamon2024fast,schuster2022many,blok2021quantum,qi2019measuring,zhang2025quantum}. 
In contrast to more typical correlation functions, out-of-time-order correlators (OTOCs) perform an unusual combination of forward and backward time evolution, making them more sensitive to the underlying model, but more difficult to implement, which has limited their use on quantum hardware. 

The main challenge for algorithms that measure OTOCs is the requirement of backward time evolution. Although this step is straightforward to implement in a digital gate-based quantum computer, it significantly increases the quantum circuit depth, making it inaccessible to NISQ and EFTQ devices. In contrast, while analog systems can more naturally simulate deep time evolution, reversing the evolution is inherently difficult, except in some special cases~\cite{liang_observation_2024}, since hardware constraints typically prevent the simple negation of relevant control parameters, as in quantum computers based on neutral atoms.

An alternative procedure based on randomized measurements has been proposed to obtain OTOCs without the need for backward time evolution. The procedure relies on two important results: the infinite temperature thermal state can be approximated via an average over matrices sampled from a Cyclic Unitary Ensemble (CUE) or a unitary $2$-design, and the OTOC operator can be broken into two parts that are each simpler to measure and whose correlations produce the OTOC value~\cite{vermersch_probing_2019}. It has been implemented in gate-based quantum computers and shown to capture the scrambling dynamics of quantum many-body systems~\cite{asaduzzaman_sachdev-ye-kitaev_2024,asaduzzaman2024quantum,joshi_probing_2022,nie_detecting_2019}. However, random unitaries are difficult to implement in a fully analog environment. To circumvent this difficulty, novel protocols have proposed using a sequence of randomized quenches to approximate unitary $n$-design properties, and implemented this technique to calculate R{\'e}nyi entropies~\cite{elben_renyi_2018,vermersch_unitary_2018,elben_randomized_2022,elben_statistical_2019}. By combining these two ideas, we develop a new protocol that can measure OTOCs with the current hardware constraints of analog quantum computers.

\begin{figure*}
    \centering
    \includegraphics[width=\textwidth]{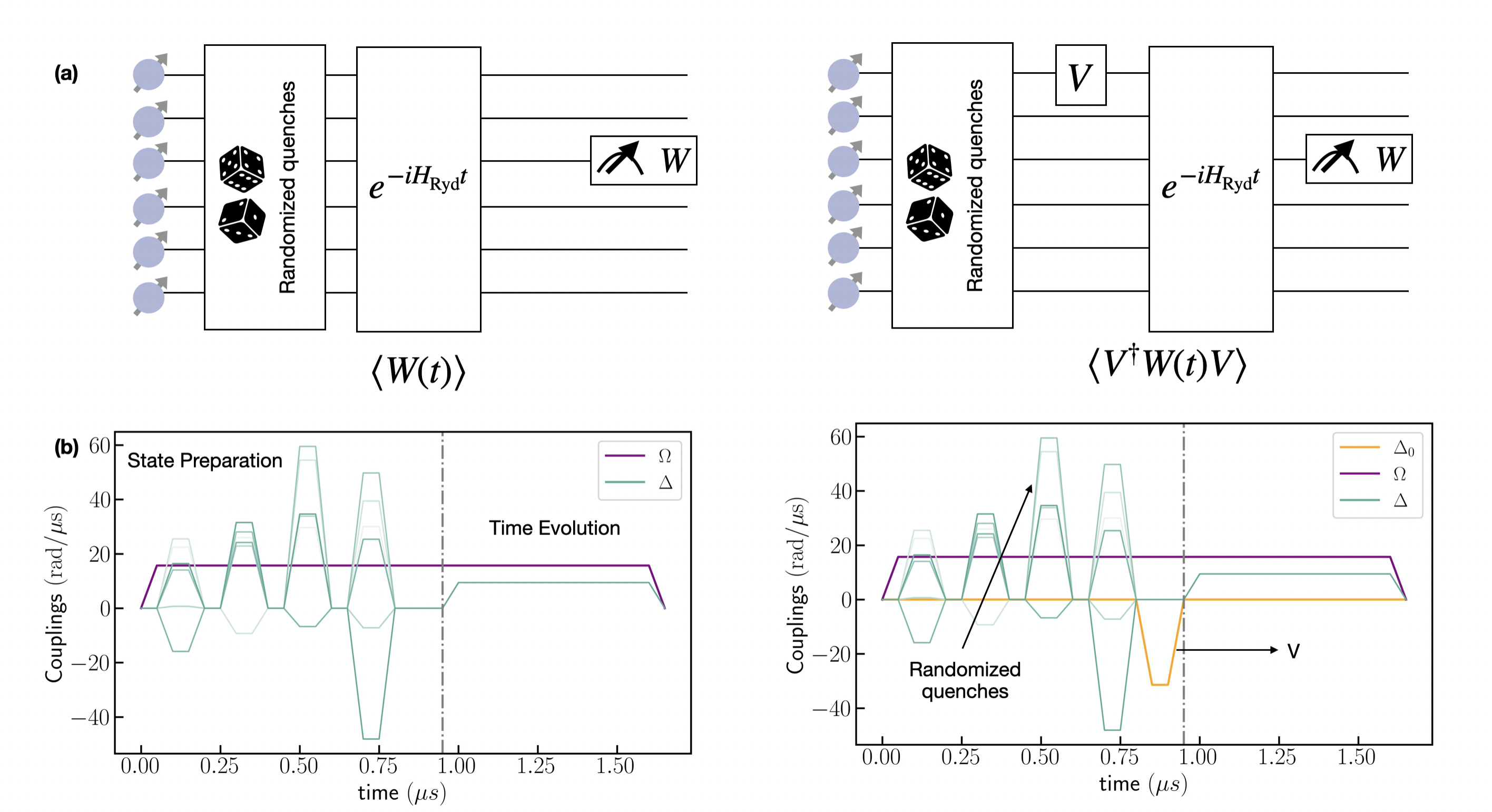}
    \caption{
    A summary of our measurement protocol. (a) The randomized analog OTOC measurement protocol, where the OTOC is calculated using the correlations between two local measurements. The first circuit diagram measures $\langle W(t)\rangle$, where $W(t) = e^{-iHt}We^{iHt}$ is the time-evolved operator in the Heisenberg picture. It implements a set of random quenches followed by analog time evolution and local measurements. The second circuit diagram measures $\langle V^\dagger W(t) V\rangle$, where $V$ is a local operator, implemented by including it before the analog time evolution. (b) The waveform schedule to implement the protocol on a Rydberg simulator. These consist of the global and local Rabi frequencies, and the detuning $\Delta$. Each green curve represents a different instance of the randomized measurements used to calculate the OTOCs.}
    \label{fig:mainfig}
\end{figure*}

In this paper, we develop and implement a randomized measurement algorithm to compute OTOCs on Aquila, the analog neutral-atom quantum computer by QuEra Computing~\cite{wurtz2023aquila}.  We show that global randomized quenches, after optimization for the control parameters of the Rydberg Hamiltonian, can approximate the properties of a unitary $2$-design needed for OTOC calculation. We then show that the algorithm reveals the information propagation dynamics of the Rydberg Hamiltonian, comparing the hardware results to tensor network calculations in the infinite temperature limit without the randomization procedure.   Our work establishes the first demonstration of a fully analog randomized OTOC measurement performed in neutral atom simulators and opens pathways for probing information propagation and quantum chaos across the rich Rydberg Hamiltonian phase diagram.  
Although we focus on neutral atom systems, our algorithm can be applied to any analog quantum computer that has control over the Hamiltonian parameters.

A graphical description of our protocol for calculating $O(t)=\langle W(t)V^\dagger W(t)V\rangle $ can be seen in Fig.~\ref{fig:mainfig}~(a), where we show the two observables of interest $\langle W(t)\rangle$ and $\langle V^\dagger W(t)V\rangle$ measured by implementing global randomized quenches and analog time evolution. In Fig.~\ref{fig:mainfig}~(b), we show how this procedure is implemented on the Aquila device by plotting the control pulses as a function of time. As indicated on the plots, our procedure can be broken into two parts, where the initial randomized quenches can be thought of as state preparation, followed by analog time evolution and measurement.

The layout of the paper is as follows. In Section~\ref{section:Methods} we introduce the details of the analog randomized measurement protocol for OTOCs and explain our implementation on Aquila. We then discuss our hardware results and compare with simulator and tensor network results in Section~\ref{section:Results}.

\begin{figure*}
    \centering
    \includegraphics[width=\textwidth]{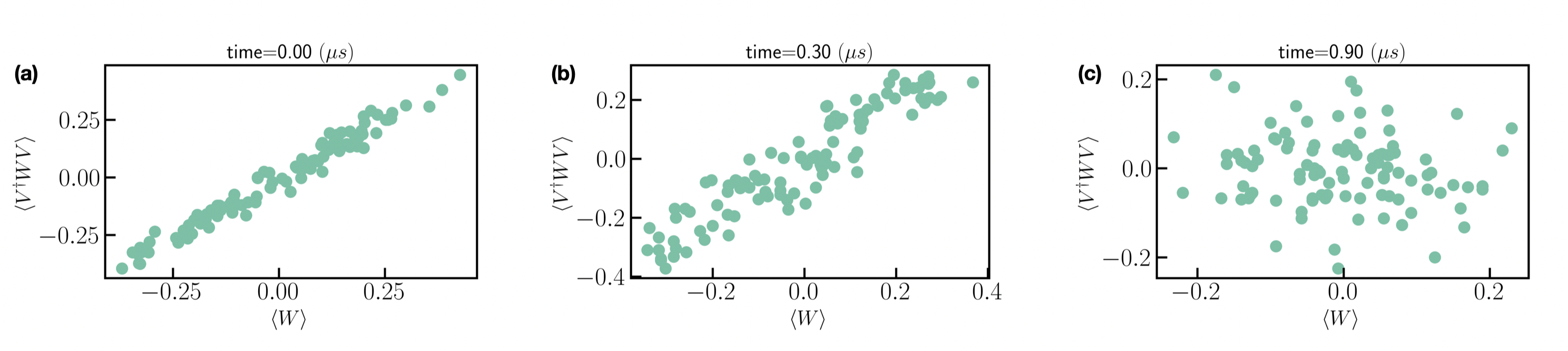}
    \caption{The decay of correlations between $\langle W(t)\rangle $ and $\langle V^\dagger W(t) V\rangle$ under analog time evolution.}
    \label{fig:scatter}
\end{figure*}

\section{Methods}\label{section:Methods}

Our goal is to develop and demonstrate a protocol to measure OTOCs throughout the Rydberg atom phase diagram, so that we can explore information propagation in the various phases. The protocol should accommodate current hardware constraints in Rydberg atom quantum simulators, notably the platform Aquila which is used here.
We will study the native Rydberg model, for which 

the Hamiltonian is
\begin{align}
   H_{\rm R} = \sum_{j} \Omega_j(t) X_j - \sum_{j} \Delta_j(t) n_j + \sum_{j<k} V_{jk} n_j n_k
\end{align}
where the sums run over the atoms, $X =\ketbra{0}{1} + \ketbra{1}{0}$, and $n = \ketbra{1}{1}$. The first term contains the Rabi drive $\Omega$, the second term contains the detuning $\Delta$, and the third term contains the Rydberg interactions in Van der Waals form, $V_{jk} \propto 1/r_{jk}^6$, where $r_{jk}$ is the distance between atoms $j$ and $k$. 
The Hamiltonian has been shown to possess a rich equilibrium phase diagram and non-equilibrium dynamical response \cite{browaeys2020many,Keesling:2018ish,darbha2024long,darbha2024false,bernien2017probing,turner2018weak,bluvstein2021controlling,ebadi2021quantum}.

Our observable of interest, OTOCs, are four-point correlation functions defined as
\begin{equation}
    O(t) = \langle W(t)V^\dagger W(t)V\rangle 
\end{equation}
where $W$ and $V$ are some selected local operators and $W(t) = e^{-iHt}We^{iHt}$ is the time-evolved operator in the Heisenberg picture. The time-evolved and time-independent operators are arranged in an out-of-time-ordered fashion.

The randomized measurement protocol for OTOCs works by factorizing the OTOC into two parts, and using the statistical correlations between these two measurements obtained from an ensemble average over a randomized set to calculate the unfactorized OTOC,
\begin{equation}
    O(t) = \overline{\langle W(t) \rangle \langle V^\dagger W(t)V \rangle}.
\end{equation}
The ensemble is generated with the application of random unitaries sampled from a CUE that approximates the infinite temperature state, so the resulting OTOCs are calculated at infinite temperature. To generate these random unitaries, we employ the procedures described in Refs.~\cite{elben_renyi_2018,vermersch_unitary_2018,elben_randomized_2022,elben_statistical_2019}, where it has been demonstrated that random quenches can approximate unitary $n$-designs. The expected behavior for our observables is that at early times, the two measurements will be highly correlated, resulting in an ensemble average close to one. As time evolves, these correlations will decay, resulting in a decrease in the statistical average. By tracking this decay, we can extract the OTOCs~\cite{vermersch_probing_2019}.

The decay of the correlations can be seen in Fig.~\ref{fig:scatter}, where we show instances of the randomized protocol for both $\langle W(t)\rangle$ and $\langle V^\dagger W(t) V\rangle$, obtained from the Bloqade simulator, at three different times. 
We see that these two measurements 
are highly correlated at $t=0.0~\mu\rm{s}$, where most of the randomized instances are clustered around the diagonal. As time evolves, the correlation between the measurements decays, which can be identified by the increased spread of the measurement points off the diagonal. By $t=0.90~\mu\rm{s}$, the observables are thoroughly uncorrelated. This shows that our implementation of the random quench procedure can successfully mimic the $2$-design properties of random unitaries needed for extracting OTOCs. 

\begin{figure}[b]
    \centering
    \includegraphics[width=0.5\textwidth]{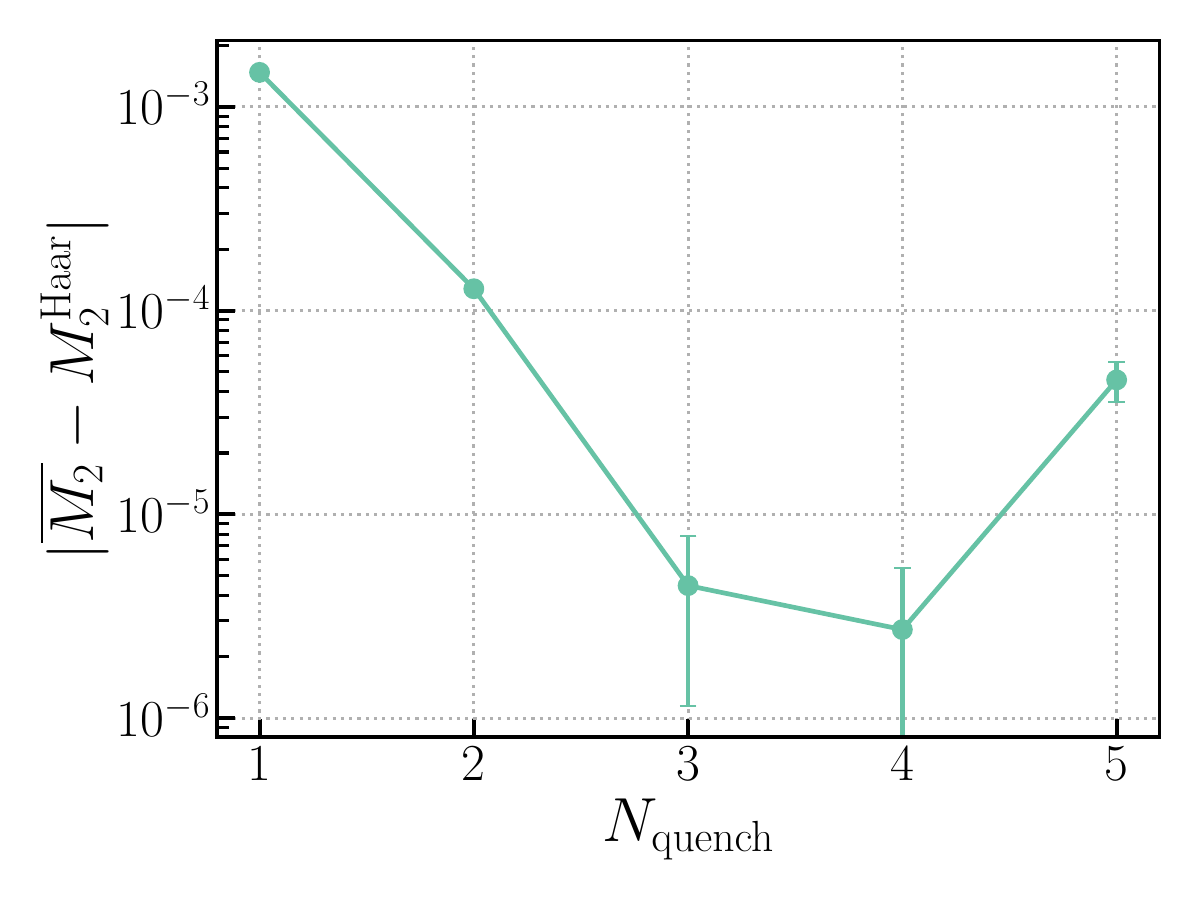}
    \caption{
    The difference in the average second moment of the probability distribution over unitaries [Eq.~\eqref{eq:M2}] computed using two methods: the quench procedure, with $N_U=200$ unitaries and $t_{\rm quench}=0.1~\mu\rm{s}$, and the Haar average $M_2^{\rm Haar}$ [Eq.~\eqref{eq:M2avgHaar}], which captures the second moment of a $2$-design. Error bars denote the standard error of the mean over random quenches.
    }
    \label{fig:convergence_to_haar}
\end{figure}

However, our protocol diverges from the implementations of Refs.~\cite{elben_renyi_2018,vermersch_unitary_2018,elben_randomized_2022,elben_statistical_2019} in two ways. Firstly, the current Aquila machine has a limited ability to address individual atoms, so we implement global quenches with randomized amplitudes. Secondly, the machine has finite coherence time and slew rates, so we also fix the duration $t_{\rm quench}$ of each quench~\cite{wurtz2023aquila}. 
Alg.~\ref{alg:otoc_short} enumerates the steps of our protocol, which was graphically represented in Fig.~\ref{fig:mainfig}.

A crucial part of our protocol is the optimization of the quench procedure
to ensure that we recover the desired $2$-design properties of a random unitary matrix~\cite{gross2007evenly,roy2009unitary}. 
The optimization involves multiple control parameters, such as the number of individual quenches, the duration of each quench, the ramp times before and after the quench, the mean $\mu/2\pi$ and variance $\sigma^2/2\pi$ of the Gaussian distribution from which the quenches are sampled, the duration of the total quench procedure, and so on. Each of these control parameters must be adjusted in accordance with the Hamiltonian parameters, namely $\Delta$,~$\Omega$,~$a$.

In light of the total coherence time $\sim 4.0~\mu\rm{s}$, we limit the quench stage to $1.0~\mu\rm{s}$, the duration and spacing of quenches to $0.1~\mu\rm{s}$, and the ramp up/down times to $0.05~\mu\rm{s}$, thereby permitting only $4-5$ quenches. We find that this number is sufficient to reliably extract OTOCs in 1D chains with up to $N=11$ atoms and atom distances $a=8.4-9.5~\mu\rm{m}$.

To demonstrate this, we examine the second moments of the probability distribution over unitaries,
\begin{equation}
    M_2(U)=\sum_{s=0}^{N-1} P_U(s)^2 \, .
    \label{eq:M2}
\end{equation}
Fig.~\ref{fig:convergence_to_haar} presents the difference between the average $\overline{M_2}$ computed from the quench procedure over $N_U$ samples and the expected Haar average
\begin{equation}
   M_2^{\rm Haar} = \frac{1+\Tr(\rho^2)}{N(N+1)} \, ,
   \label{eq:M2avgHaar}
\end{equation}  
as a function of the number of quenches $N_{\rm quench}$. 
 
The figure shows that our random quenches approximate the second moment of a $2$-design  even under the hardware restrictions. The choice of $N_{\rm quench}=4$ nearly optimally captures the $2$-design properties needed for the purposes of OTOC extraction at our fixed quench duration. 
The remaining parameters, i.e. the mean and variance of the Gaussian distribution for the quench parameters, are optimized to obtain the most accurate observed lightcone. 

\begin{algorithm}[t]
\caption{Analog randomized OTOC measurement on a Rydberg array}
\label{alg:otoc_short}
\begin{algorithmic}[1]
\FOR{$u=1,\dots,N_U$}
    \STATE Prepare $\ket{\psi_0}=\ket{0}^{\otimes N}$
    \STATE Apply $N_{\rm quench}$ global random quenches
    \STATE Evolve under $H_{\rm Rydberg}$ for time $t$
    \STATE Measure Rydberg density $n$ to estimate $\langle W(t)\rangle_u$

    \STATE Prepare $\ket{\psi_0}$ and apply the \emph{same} random quenches
    \STATE Apply local $V$ (detuning pulse)
    \STATE Evolve under $H_{\rm Rydberg}$ for time $t$
    \STATE Measure $n$ to estimate $\langle V^\dagger W(t)V\rangle_u$
\ENDFOR
\STATE Compute OTOC:
\begin{equation*}
    O(t) = \frac{1}{N_U} \sum_{u=1}^{N_U} \langle W(t)\rangle_u \langle V^\dagger W(t)V\rangle_u
\end{equation*}
\end{algorithmic}
\end{algorithm}

\begin{figure*}
    \centering
    \includegraphics[width=\textwidth]{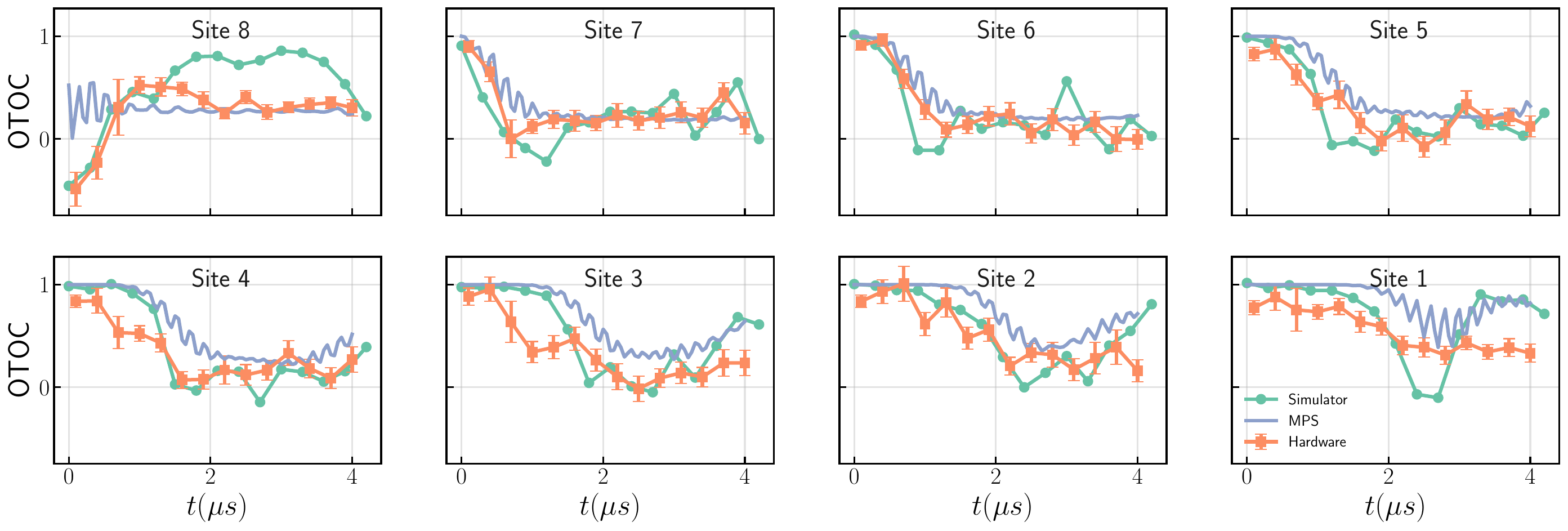}
    \caption{Comparison between different methods of measuring OTOCs.}
    \label{fig:lineplot_OTOCs}
\end{figure*}

Since we have shown that global quenches accurately approximate $2$-design properties, we can turn to the rest of the protocol. We initialize each atom in its ground state, then perform random quenches followed by analog time evolution. 
We choose  $W_i=n_i$, and apply the $V$ operator on the last atom of the chain as a local detuning pulse, which results in $O_{ij}(t)=\langle n_i(t)n_jn_i(t)n_j\rangle$. 
According to our protocol, this procedure is then repeated twice for the two observables we are interested in, which are $\langle n_i(t) \rangle $ and $\langle n_j^\dagger n_i(t) n_j\rangle$. 
After analog time evolution, we measure the Rydberg density for both observables. We can repeat this procedure $N_U$ times to build our ensemble average. Each random sample is then measured $N_S$ times to extract statistics. The overall shot cost of the protocol then comes to $N_{\rm total} = 2 N_U N_S$. 

Since $\langle W(t)\rangle \sim 1/\sqrt{2^{N_{\rm total}}}$, as derived in previous work~\cite{vermersch_probing_2019}, the scaling $\epsilon\sim 2^{N_{\rm total}}$ gives the number of shots required to obtain an OTOC measurement below a given error $\epsilon$. 
This exponential scaling makes the shot count the
most costly item in the protocol. 
After we measure both observables to the desired precision, OTOCs can be calculated via the ensemble average over the randomized quenches.

\section{Results}\label{section:Results}

To demonstrate our protocol, we examine a fiducial point in the Rydberg phase space: $\Omega/2\pi=2.5~\rm{MHz}$ and $\Delta/2\pi=1.5~\rm{MHz}$ for the Rabi and detuning terms, and $a=9.5~\mu\rm{m}$ for the atom distance in the chain. 
This results in a blockade radius $R_b = (C_6/\Omega)^{1/6}\sim 0.86 a$, so that all our atoms are outside but near the Rydberg blockade. These parameters demonstrate our protocol in a simple limit where the Rydberg Hamiltonian approaches a nearest-neighbor interacting model.  However, this choice is arbitrary and should not affect our protocol, so we also probe other parameters to check consistency.

\begin{figure}[b]
    \centering
    \includegraphics[width=0.5\textwidth]{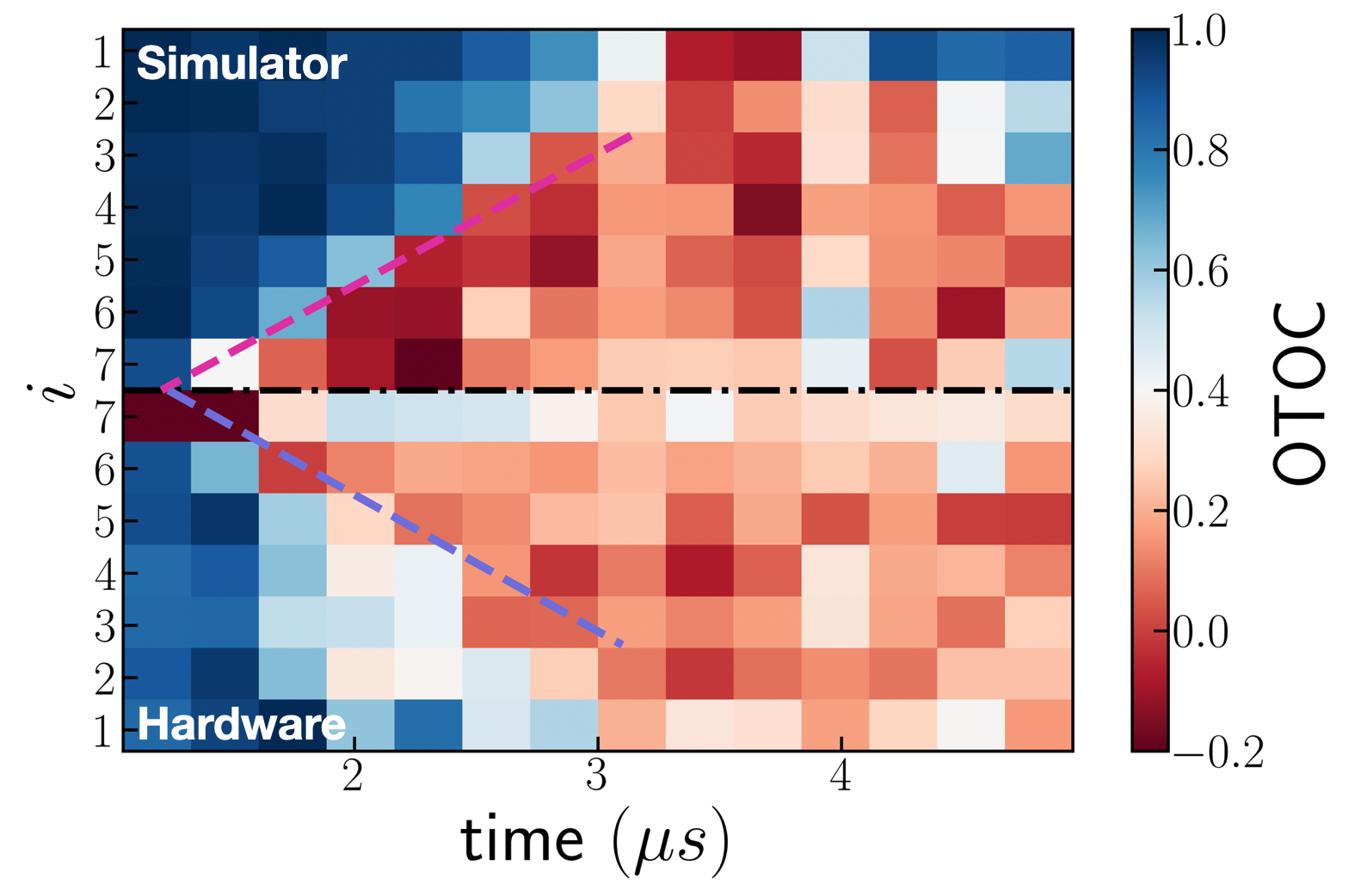}
    \caption{
    Comparison between the simulator (top) and hardware OTOC measurements (bottom) for atom distance $a = 9.5~\mu\rm{m}$. 
    Dashed lines indicate the edge of the lightcone obtained by fitting the data to a linear function
    }
    \label{fig:heatmap_comparison}
\end{figure}

\begin{figure*}[t]
    \includegraphics[width=0.95\textwidth]{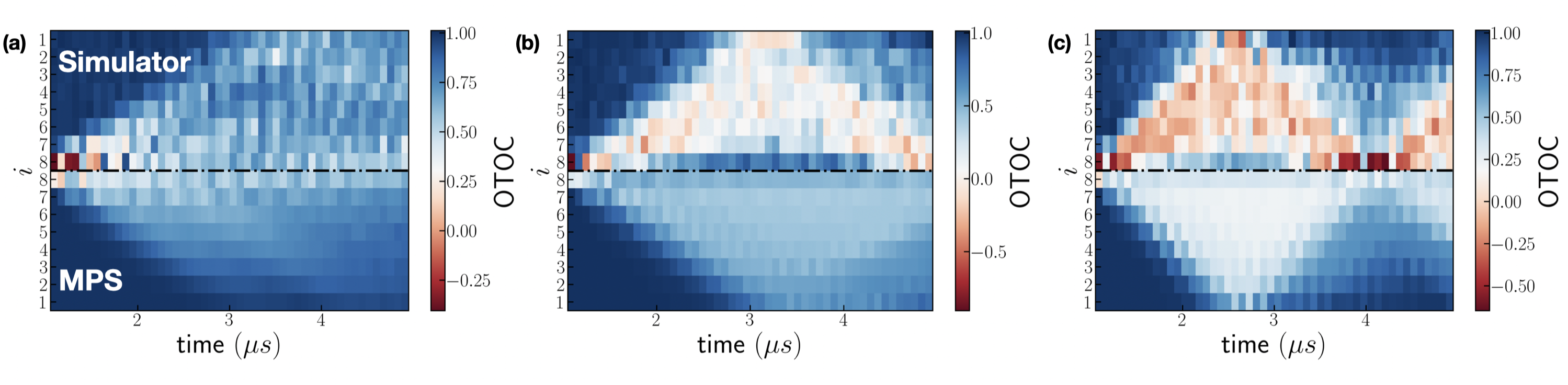}
    \caption{
    Stability of the optimized quench protocol for Hamiltonian parameters that deviate from our main case $\Delta/2\pi=1.5~\rm{MHz}$, $\Omega/2\pi=2.5~\rm{MHz}$, $a=9.5~\mu\rm{m}$, namely (a) higher detuning $\Delta/2\pi=2.5~\rm{MHz}$, (b) lower Rabi frequency $\Omega/2\pi=2.0~\rm{MHz}$, and (c) smaller atom distance $a=9.0~\mu\rm{m}$. OTOC results are shown from the Bloqade simulator in the top half and tensor network simulations in the bottom half, with time on the $x$-axis and the Rydberg atom index $i$ on the $y$-axis.
    }
    \label{fig:heatmaps}
\end{figure*}

We ran the protocol described in the previous section on the Aquila machine by QuEra Computing, and locally post-processed the hardware data to extract the OTOCs. To check the hardware results, we also ran the same protocol on the Bloqade simulator and a noisy simulation\footnote{Noisy simulations used QuTiP~\cite{qutip5}}, both of which are state vector simulations. We also conducted matrix product state (MPS) tensor network simulations\footnote{MPS calculations used ITensor~\cite{Fishman:2020gel}} corresponding to true infinite temperature OTOCs~\cite{RevModPhys.77.259,xu2020accessing} to benchmark the success of the protocol in measuring infinite temperature OTOCS.  

In Fig.~\ref{fig:lineplot_OTOCs}, we plot OTOC results obtained from the hardware together with the Bloqade simulator and tensor network simulations. On the $8$th site, where we implemented a large local detuning pulse, the MPS results closely match the hardware measurements for $t \gtrsim 1~\mu\rm{s}$. In contrast, the simulator diverges from the hardware measurements, indicating that it fails to capture the relaxation dynamics on this site. At the neighboring $7$th site without the local detuning, the three methods yield close agreement. Traversing the chain, the MPS results increasingly diverge from the hardware data, since the signal requires more travel time. 

To understand the discrepancy between the simulator and hardware results, we carried out a noisy simulation by introducing a homogeneous depolarizing channel, Rydberg to ground state decay, laser detuning, and atom noise. We saw that in the absence of the noise, the correlations between $\langle W(t)\rangle$ and $\langle V^\dagger W(t)V\rangle$ do not decay as expected, and these two measurements start correlating again, which results in an unexpected increase in the OTOC. However, in the presence of noise, correlations between these two measurements decay as expected, and our simulation and hardware results begin to agree with the MPS results, as seen in Fig.~\ref{fig:lineplot_OTOCs}. Detailed plots of the noisy simulations are provided in App.~\ref{App:Noisy_sim} and plots  of the correlations between the two measurements are provided in App.~\ref{App:Full_data}. The fact that the noisy simulation agrees more closely with the expected results than the Bloqade simulator shows that hardware noise actually helps by introducing additional uncertainty in the quench procedure, alleviating some of the effects of our restrictions on the protocol, and gives a better approximation to $2$-design unitaries used for extracting OTOCs.

The structure of the lightcone can be observed explicitly by collecting the site OTOCs into a heatmap, as shown in Fig.~\ref{fig:heatmap_comparison} for the hardware and simulator data. We omit the $8$th site due to the discrepancy between the simulator and the rest of the methods. For the other sites, the hardware and simulator results show good agreement up to $4.0~\mu\rm{s}$. This timescale is approximately the atom dephasing time on Aquila, resulting in the degradation of the OTOC signal at later times, and leading to poorer agreement farther away from the initial perturbation. The signal travels from atom $i = 8 \rightarrow 1$ in around $3.0-3.5~\mu\rm{s}$. It reflects back at around $4.0~\mu\rm{s}$ on the simulator, a behavior not observed on the hardware due to the atom decoherence. 

The rate of information spreading can be extracted from the shape of the lightcone. Since the atoms have a large separation, long-range interactions can be ignored, and the signal should produce a linear lightcone in accordance with the Lieb-Robinson bounds~\cite{lieb2004finite}, a prediction corroborated in the OTOC measurements in Fig.~\ref{fig:heatmap_comparison}. The speed of information propagation in this system can be obtained by fitting the lightcone to a straight line and extracting the slope.  The extracted slopes are $m_{\rm Simulator}=0.32\pm0.02$, $m_{\rm Hardware}=0.31 \pm 0.03$, and $m_{\rm MPS}=0.33\pm0.01$, all of which are consistent with each other within their respective uncertainties. 
In systems with different assumptions, the lightcone can have a different functional form, ranging from linear to exponential, due to different information propagation and scrambling behavior~\cite{sekino_fast_2008,asaduzzaman2024quantum,yuan_quantum_2022,chen2019quantum,deng2017logarithmic}.

So far, we have demonstrated the success of our measurement protocol for one specific set of parameters. However, as stated earlier, our method works at any point in the phase space. To demonstrate this, we simulate our measurement at several other points. 
One major obstacle is the need to re-optimize the quench parameters for phase space points far from the fiducial one. 
The most important parameters that affect the quench procedure are the distance between atoms, $a$, and the values for the Rabi, $\Omega$, and detuning, $\Delta$, terms.

In Fig.~\ref{fig:heatmaps} we show three instances of the OTOC measurement protocol in the vicinity of the fiducial Hamiltonian parameters $a=9.5~\mu\rm{m}$, $\Delta/2\pi=1.5~\rm{MHz}$, and $\Omega/2\pi=2.5~\rm{MHz}$, and compare the results obtained from the Bloqade simulator (top half) to the those from the MPS calculation in the infinite temperature limit (bottom half). 
In Fig.~\ref{fig:heatmaps}(a), we increase the detuning term to $\Delta/2\pi=2.5~\rm{MHz}$, yielding a small change that does not require the quench procedure to be re-optimized. The lightcone structure can be reliably recovered, and the calculated OTOCs agree well between the two methods. 
In Fig.~\ref{fig:heatmaps}(b), we change the Rabi term to $\Omega/2\pi =2.0~\rm{MHz}$, which has a much stronger effect on the OTOC values than changing the detuning. Even though the shapes of the lightcones obtained from both methods are in good agreement, the actual values of the OTOCs differ. 
Finally, in Fig.~\ref{fig:heatmaps}(c), we modify the distance between the Rydberg atoms to $a=9.0~\mu\rm{m}$. This results in the biggest difference between the simulator and MPS methods: while the shape of the lightcone is still similar in both methods, the values of the OTOC measurement are very different, which indicates the need for re-optimization of the quench procedure. 

Since the atom distance and Rabi term control most of the dynamics of the Rydberg system, we expect large discrepancies between the two results when changing those two parameters. However, despite large changes in the OTOC values, we see that it is still possible to extract the structure of the lightcone in the vicinity of the original optimized parameters.

\begin{figure}
    \centering
    \includegraphics[scale=0.2]{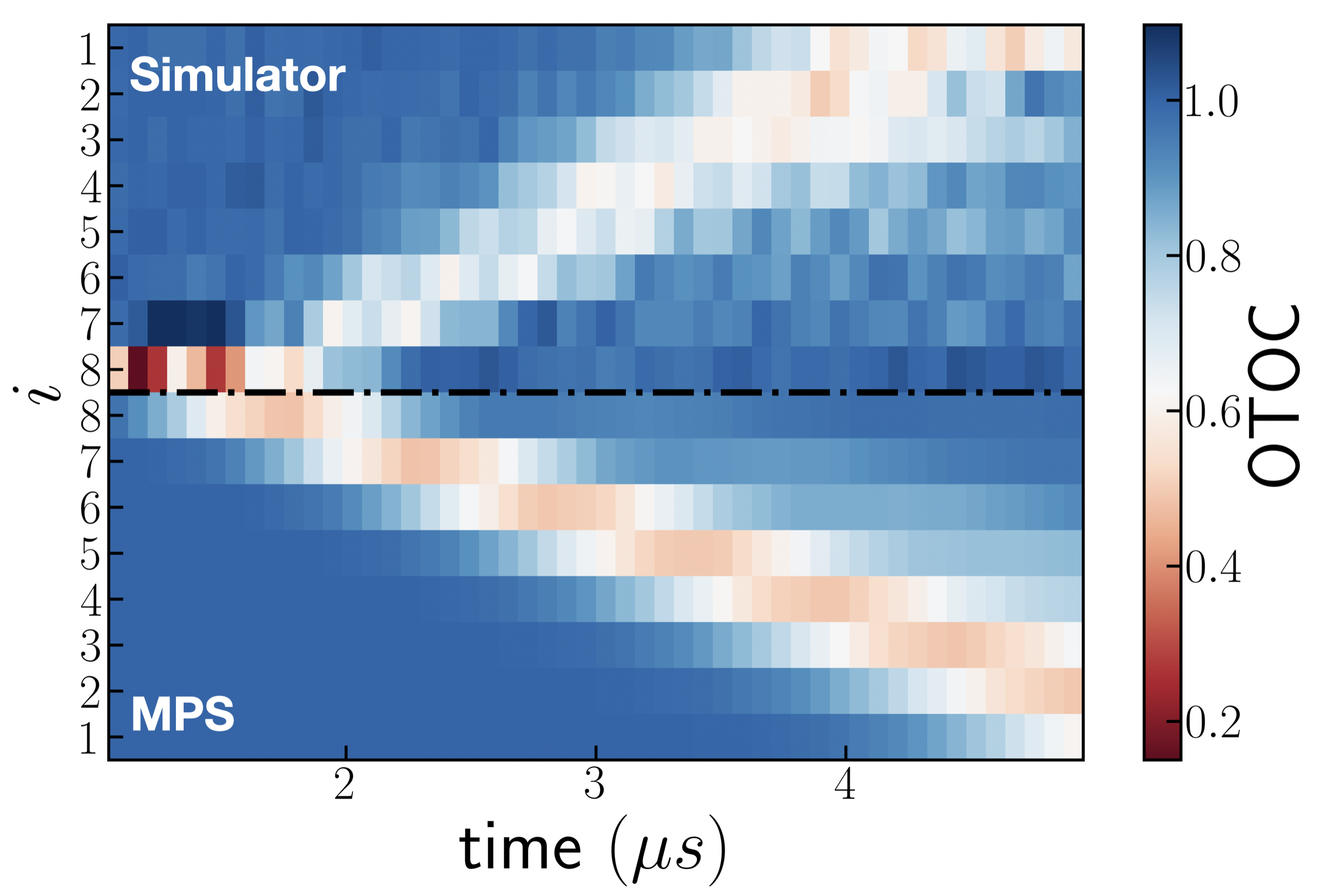}
    \caption{
    Comparison between the simulator and hardware OTOC measurements for atom distance $a = 8.4~\mu\rm{m}$. The top half shows the simulator data and the bottom half shows MPS tensor network simulations for OTOCs, with time on the $x$-axis and the Rydberg atom index $i$ on the $y$-axis.
    }
    \label{fig:heatmap_8.4}
\end{figure}

As a final example, we demonstrate the applicability of our measurement protocol for a completely unrelated point in the phase space where the dynamics are highly constrained due to the blockade mechanism. For this example, we set the atom distance to $a=8.4~\mu\rm{m}$ and the  Rabi and detuning terms to $\Delta/2\pi=2.5~\rm{MHz}$ and $\Omega/2\pi=0.4~\rm{MHz}$ respectively. We re-optimize the quench procedure and find that fixing the mean of the Gaussian distribution to $\mu/2\pi=0.75~\rm{MHz}$ gives us a good resolution for the lightcone. In Fig.~\ref{fig:heatmap_8.4}, we plot the results obtained from the Bloqade simulator and MPS simulations of the OTOCs. Due to constrained dynamics, the lightcone is shallow here, where the space under the initial propagation line remains close to $1.0$. This demonstrates that with careful optimization, our method can extract the information propagation dynamics in the different phases of the Rydberg Hamiltonian.

\section{Conclusions}

In this paper, we developed and implemented a fully analog randomized measurement protocol for measuring OTOCs on neutral atom platforms. 
Our protocol relies on approximating the infinite temperature state by applying random unitaries sampled from a CUE.  These random unitaries are then approximated by a sequence of global randomized quenches on the analog quantum computer.  Once we obtain a good approximation  of CUE unitaries, infinite temperature OTOCs are calculated  from the ensemble average of  the statistical correlations between two measurements given as $\langle W(t)\rangle $ and $\langle V^\dagger W(t) V\rangle $  over different random quench instances.  This quench procedure needs to be fine-tuned for the parameters of the Hamiltonian; however, we showed that within the vicinity of the original Hamiltonian parameters, a previously optimized quench protocol can still be used to study the lightcone of information propagation. 

We demonstrated this protocol by measuring OTOCs on the neutral atom quantum computer Aquila by QuEra Computing, and showed that it can reliably extract the lightcone and the temporal behavior of OTOCs. We compared our experimental results to both the Bloqade emulator tensor network simulations and noisy simulations, and found close agreement. For the site where the local detuning pulse implements the $V$ operation, hardware noise actually helps our quench procedure by providing extra randomization, allowing the correlations between $\langle W(t)\rangle $ and $\langle V^\dagger W(t) V\rangle $ to decay as expected. 

Since our method doesn't require any time inversion, it is implementable for any point in the Rydberg phase space, unlike some methods that require the Hamiltonian to be in the PXP limit to achieve time reversal~\cite{liang_observation_2024}. This paves the way for studying information propagation and quantum chaos in the expansive regimes of the Rydberg Hamiltonian and exploring the effects of the Rydberg blockade mechanism on information propagation.  Our method is also implementable in different kinds of analog quantum computers as long as one has access to Hamiltonian parameters. 

The largest bottleneck in this protocol is the optimization procedure for the random quenches and the number of shots needed to extract the OTOCs. The optimization can be improved greatly by implementing a reinforcement learning schemes~\cite{Bukov:2026gbf,bukov2018reinforcement,niu2019universal} to learn pulse sequences that approximate $2$-design properties without manual scanning, which we plan to implement in future work. 

Our algorithm can be also be applied in optical lattices, which capture the Fermi-Hubbard Hamiltonian.~\cite{esslinger2010fermi,dutta2015non,schafer2020tools,xu2025neutral} In this case, one can calculate OTOCs in very large systems, pushing the limits of classical calculations.

\section*{Acknowledgments}
AFK and  GT acknowledge financial support from the National
Science Foundation under award No. PHY-2325080: PIF: Software-Tailored Architecture for Quantum Co-Design. This research was supported by the U.S. Department of Energy (DOE) under Contract No. DE-AC02-05CH11231, through the National Energy Research Scientific Computing Center (NERSC), an Office of Science User Facility located at Lawrence Berkeley National Laboratory.  This research used resources of the National Energy Research Scientific Computing Center (NERSC), a Department of Energy Office of Science User Facility under Contract No. DE-AC02-05CH11231 using NERSC award for QCAN Project DDR-ERCAP0033861. We acknowledge the computing resources provided by North Carolina State University High Performance Computing Services Core Facility (RRID:${\rm SCR\_}022168$).

\bibliography{QuEra_OTOC}

\clearpage

\onecolumngrid

\appendix

\section{Details about Noisy Simulation}\label{App:Noisy_sim}

\begin{figure}[h]
    \centering
    \includegraphics[scale=0.6]{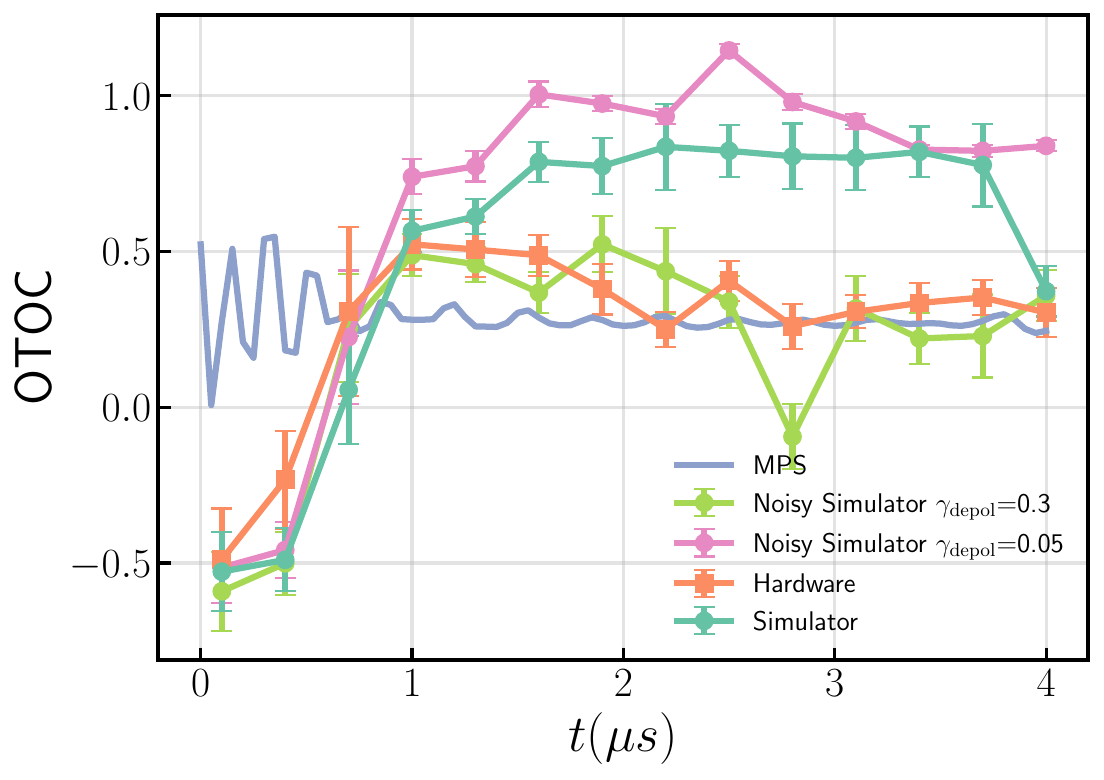}
    \caption{
    Comparison between the different levels of noisy simulator, noiseless simulator,hardware and MPS OTOC measurements for site 8. 
    }
    \label{fig:noisy_sim}
\end{figure}

To obtain the noisy results, we simulated the quench protocol with the Lindblad master equation by introducing the following jump operators,
\begin{align}
    c_{\rm depol} &= \sqrt{\gamma_{\rm depol}}(X+Y+Z) \\ 
    c_{\rm rg} &= \sqrt{\gamma_{\rm rg}}(|1\rangle\langle 0|)
\end{align}
where $X,Y,Z$ are the usual Pauli operators and  $\gamma_{\rm depol}=0.2,0.05$ and $\gamma_{\rm rg}=0.03$. We also included laser amplitude noise for both detuning, $\delta\Delta /2\pi=0.18~\rm{MHz}$, and Rabi drives, $\delta \Omega/\Omega=0.018$,  as well as noise in the atom distance, $\delta a = 0.05 ~\mu{\rm m}$. For the site with a local detuning, we increased the noise parameter two-fold to capture the effects of the local detuning, introducing extra sources of errors. 
We solved the master equation using a Monte Carlo method with $600-700$ trajectories, implemented using the QuTiP software library~\cite{qutip5}. 
Fig.~\ref{fig:noisy_sim} shows the OTOC measurement from the noisy simulations. Increasing the noise level of the depolarizing channel to $\gamma_{\rm depol}=0.2$ from $\gamma_{\rm depol}=0.05$ increases the agreement between the simulator and hardware results. 

\section{for Obtaining OTOCs}
\label{App:Full_data}

\begin{figure}[t]
    \centering
    \includegraphics[height=0.42\textheight, width=\textwidth]{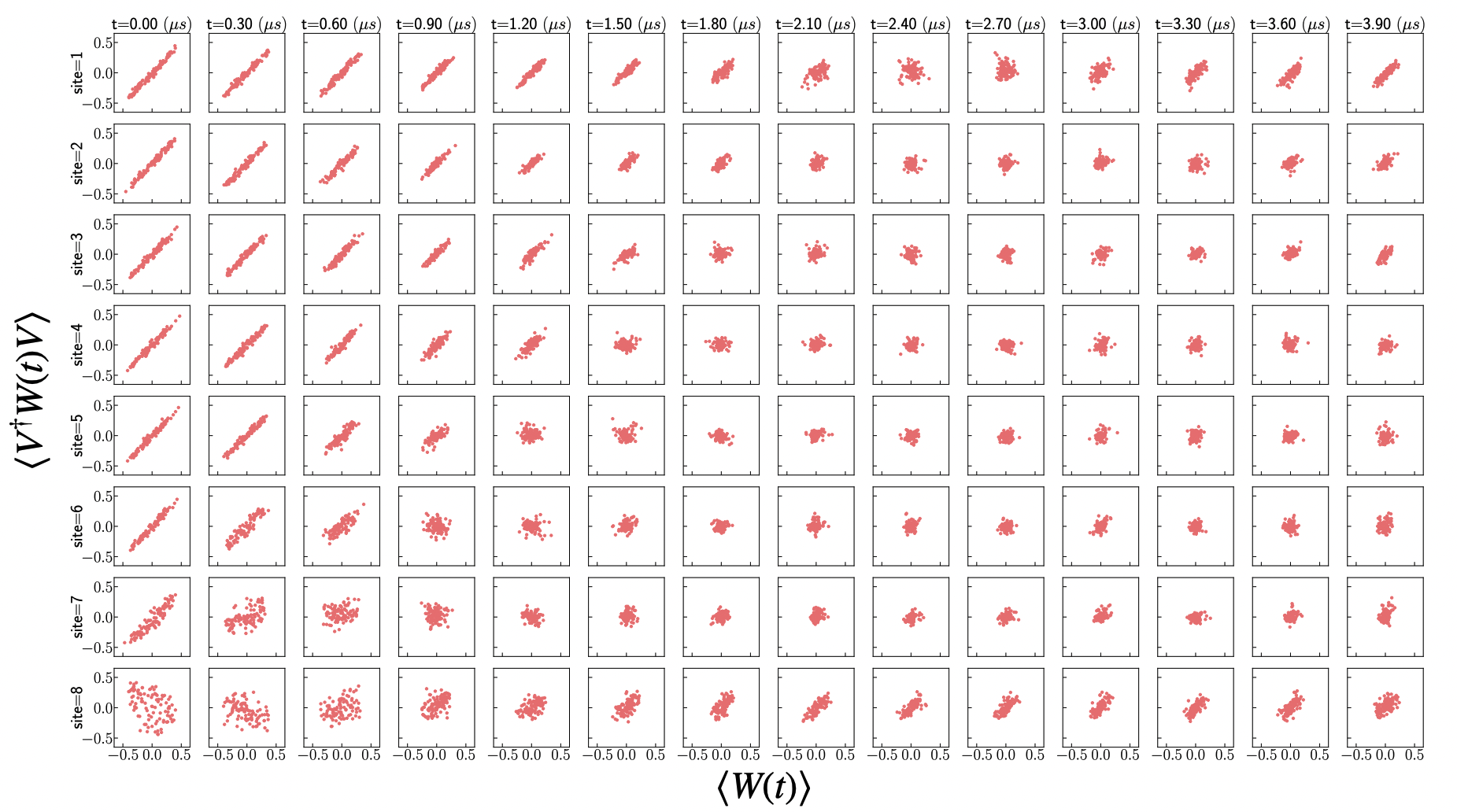}
    \caption{Simulator results for $\langle n_i(t)\rangle$ and $\langle n_8^\dagger n_i(t) n_8\rangle$. Each dot represents a single randomized instance that is repeated $N_s$ times.}
    \label{fig:full_bloqade}
\end{figure}

\begin{figure}[b]
    \centering
    \includegraphics[height=0.42\textheight, width=\textwidth]{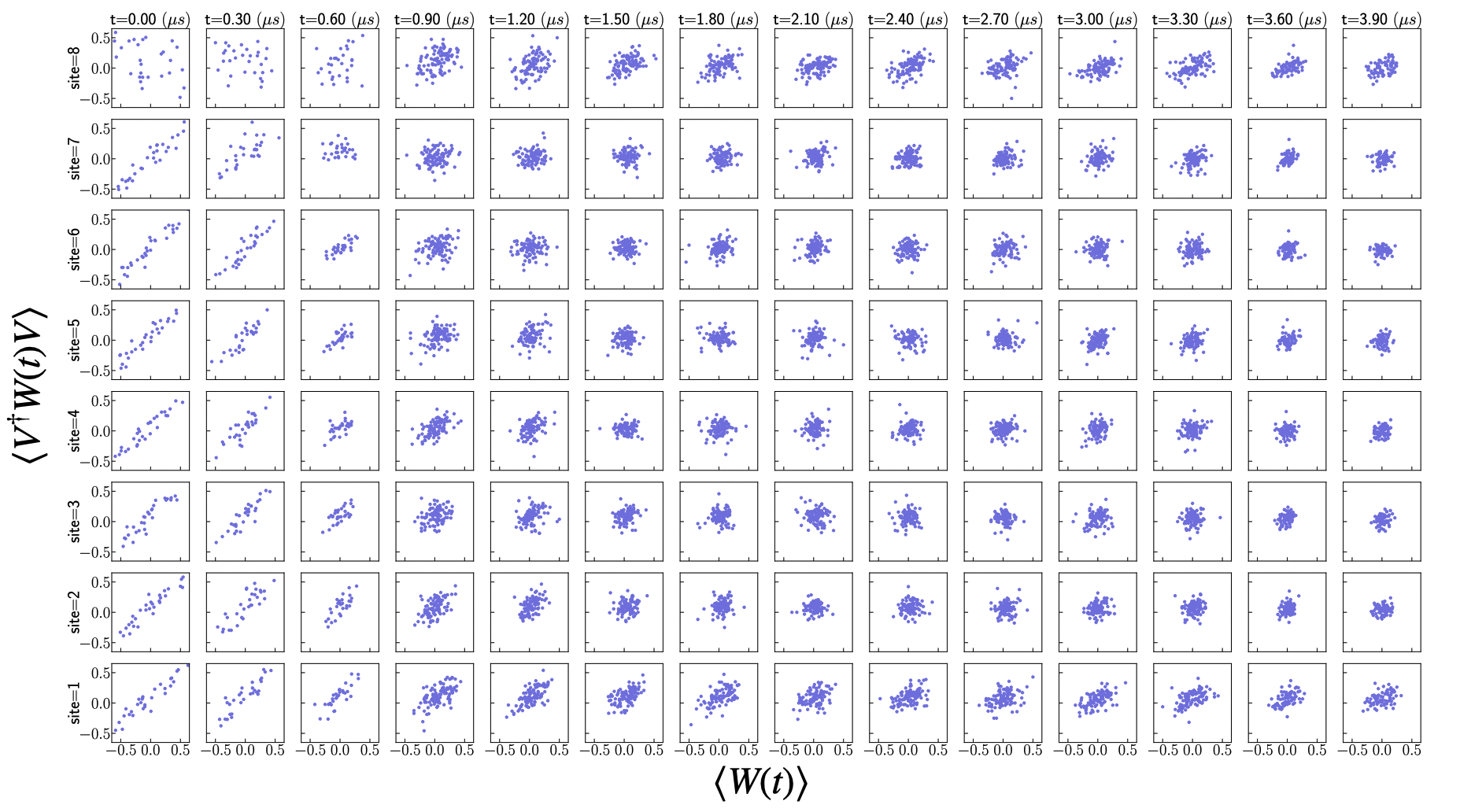}
    \caption{Hardware results for $\langle n_i(t)\rangle$ and $\langle n_8^\dagger n_i(t) n_8\rangle$. Each dot represents a single randomized instance that is repeated $N_s$  times.}
    \label{fig:full_hardware}
\end{figure}

In this appendix, we provide the results for the measurement $\langle n_i(t)\rangle$ and $\langle n_8^\dagger n_i(t) n_8\rangle$. In Fig.~\ref{fig:full_bloqade}, we show the output from the Bloqade simulator. We can see that site $8$ starts to develop some correlations between the two measurements again around $t\sim 1.80~\mu\rm{s}$, which results in an increase in the OTOC extracted from the statistical correlations between these two measurements. In Fig.~\ref{fig:full_hardware} we show the data obtained from Aquila, the effects of the hardware noise compared to the simulator can be seen from the enhanced spreading of measurements compared to the simulator. For our constrained protocol, the noise actually helped us in approximating the needed $2$-design properties. These plots have the same layout as in Fig.~\ref{fig:heatmap_comparison} to keep them aligned with the propagation pattern there. The lightcone of information propagation can also be seen in Figs.~\ref{fig:full_bloqade}~and~\ref{fig:full_hardware} by tracking the change in the distribution from a diagonal spread to a circular one.

\newpage

\end{document}